\documentstyle[preprint, aps]{revtex}
\tightenlines
\begin{document}
\title{Wigner distributions for non Abelian finite groups of odd order}
\author{N. Mukunda\thanks{email: nmukunda@cts.iisc.ernet.in}}
\address{Centre for Theoretical
Studies, Indian Institute of Science, {Bangalore~560~012,} India\\
and\\
Jawaharlal Nehru Centre for Advanced Scientific Research,
 Jakkur, Bangalore~560~064, India}
\author{S. Chaturvedi\thanks{e-mail: scsp@uohyd.ernet.in}}
\address{ School of Physics, University of Hyderabad, Hyderabad 500 046,
India}
\author{ R.Simon\thanks{email: simon@imsc.res.in}}
\address{The Institute of
 Mathematical Sciences, C. I. T. Campus, Chennai 600 113, India}
 \date{\today}
 \maketitle
\begin{abstract}
Wigner distributions for quantum mechanical systems whose configuration space
is a finite group of odd order are defined so that they  correctly reproduce
the marginals and have desirable transformation properties under left and right
translations. While for the Abelian case we recover known results,
though from a different perspective, for the non Abelian case our
results appear to be new.
\end{abstract}
\newpage
The notion of a Wigner distribution, introduced about seventy years ago in
the context of quantum mechanical systems with ${\cal R}^n$ as the
configuration space\cite{1}, has played a dominant role in various areas in
physics both as a bridge between classical and quantum mechanics as well as
 a useful computational tool\cite{2}. In the last twenty years there have
been several attempts to extend the definition of the Wigner distribution
beyond the Cartesian case ${\cal R}^n$ \cite{3}-\cite{12}.
In a recent work\cite{13} in this general direction, a notion of a Wigner
distribution was developed from first principles for the case when the
configuration space is a compact Lie group $G$. A detailed analysis of
the quantum kinematics of such systems  revealed that the transition 
from ${\cal R}^n$ to an arbitrary compact Lie Group $G$ entails several 
extensions or modifications of the familiar Wigner distribution formalism 
in the Cartesian case. Some novel features of this formalism are listed below:

\begin{itemize}
\item  The role of the continuous momenta in the Cartesian case is  played by 
the discrete labels $JMN$ of the unitary irreducible representation matrices
${\cal D}^{J}_{MN}(g)$ of the group $G$. Here $J$ labels the irreducible
representations of $G$ and $MN$ the rows and the columns.

\item  If one insists on the recovery of the marginal distributions and correct
  transformation properties under left and right translations, the
  Wigner distribution associated with a normalised wavefunction $\psi(g)$
on the compact Lie group $G$ no longer turns out to be just a function
$W(g;JMN)$ of the `position coordinates' $g$ and the `momentum coordinates'
$JMN$ alone but rather a more elaborate object $W(g;JMN~M^\prime N^\prime)$ :
\begin{eqnarray}\label{1}
W(g; JMN\;M^{\prime}N^{\prime}) &=& N_J\int\limits_G dg^{\prime}
\int\limits_G dg^{\prime\prime}\; \delta\left( g^{-1}
s(g^{\prime},g^{\prime\prime})\right)\nonumber\\ &&{\cal D}^J_{MN}
(g^{\prime})\;\psi(g^{\prime})^*\; {\cal
D}^J_{M^{\prime}N^{\prime}}(g^{\prime\prime})^*\;
\psi(g^{\prime\prime}).
\end{eqnarray}
Here $N_J$ is the dimension of the unitary irreducible representation
$J$ and $dg$ is the normalised translation invariant volume element on $G$. 

In the case of Abelian groups, simplifications occur owing to the
fact that all irreducible representations are one dimensional
making the labels $M~N~M^\prime~N^\prime$ redundant with the
result that the Wigner distribution
$W(g;JMN\;M^{\prime}N^{\prime})$ can simply be written as
$W(g,J)$, a function of just the coordinates `$g$' and the momenta
`$J$' as in the cases of ${\cal R}^n$ and ${\cal S}^1$.

\item  A key ingredient in the construction of the Wigner function above is
  the notion of the `mid point' $s(g,~g^\prime)$ of two group elements $g$
and $g^\prime$. (Properties of $s(g,~g^\prime)$ are given in detail later).
In the compact Lie group case, this object can be computed using the geodesics
on the manifold of $G$.

\item  The form of the Wigner distribution in $(\ref{1})$ has a wider
range of applicability than just compact Lie groups as was
demonstrated in \cite{13} by recovering the known results for the
non compact case ${\cal R}^n$ and ${\cal S}^1$ \cite{14}. In these
cases, owing to the Abelian nature of the groups involved, though
the geodesic construction is not available for computing
$s(g,~g^\prime)$ , it does turn out to be possible to find
$s(g,~g^\prime)$ satisfying the desired conditions and hence the
known Wigner distributions appropriate to these cases.

\end{itemize}
The purpose of  the present work is to show that the structure of the
 Wigner distribution given above, suitably adapted to finite groups,
leads to a satisfactory definition of Wigner distributions for all finite
groups of odd order both Abelian and non Abelian.

Let ${\cal H}$ denote the Hilbert space of complex valued functions on a
finite group $G$ of order $|G|$ and let $\psi(g), ~g\in G$ denote a
normalised `position-space' wavefunction:
\begin{equation}
||\psi||^2 \equiv \sum_{g\in G} |\psi(g)|^2 =1.
\end{equation}
Let ${\cal D}^{J}_{MN}(g)$ denote the unitary irreducible
representation matrices of $G$ with $J$ labelling the irreducible
representation and $MN$ the rows and columns respectively. These
matrices satisfy the following representation, orthogonality and
completeness properties:
\begin{eqnarray} 
\sum\limits_{M^{\prime}} {\cal D}^J_{MM^{\prime}}
(g^{\prime}) {\cal D}^J_{M^{\prime}N} (g) &=&
{\cal D}^J_{MN}(g^{\prime}g) ,\label{2a}\\
\frac{1}{|G|}\sum_{g\in G}\;{\cal D}^{J^{\prime}}_{M^{\prime}N^{\prime}}
(g)^*{\cal D}^J_{MN}(g) &=& \delta_{J^{\prime}J}
\delta_{M^{\prime}M}\delta_{N^{\prime}N} /N_J ,\label{2b}\\
\frac{1}{|G|}\sum\limits_{JMN} N_J {\cal D}^J_{MN} (g) {\cal
D}^J_{MN}(g^{\prime})^* &=& \delta_{g, g^{\prime}} .\label{2c}
\end{eqnarray}
With the help of the matrices ${\cal D}^{J}_{MN}(g)$ we define the
`Fourier transform' $\psi_{JMN}$ of $\psi(g)$ :
\begin{eqnarray}  \psi_{JMN}& =& \sqrt{\frac{N_J}{|G|}}
\sum_{g\in G}\;{\cal D}^J_{MN}
(g)^* \psi(g) ,\nonumber\\
\sum_{JMN}|\psi_{JMN}|^2 &=& \sum_{g\in G}|\psi(g)|^2,
\end{eqnarray}
yielding the `momentum-space' wavefunction $\psi_{JMN}$.

Following \cite{13}, we associate with $\psi(g)$ the Wigner distribution
$W(g;JMN\;M^{\prime}N^{\prime})$ as follows:
\begin{eqnarray}\label{10}
W(g; JMN\;M^{\prime}N^{\prime}) &=&
\frac{N_J}{|G|}\sum_{g^{\prime}\in G} \sum_{g^{\prime\prime}\in
G}\; \delta_{ g, s(g^{\prime},g^{\prime\prime})}\nonumber\\
&&{\cal D}^J_{MN} (g^{\prime})\;\psi(g^{\prime})^*\; {\cal
D}^J_{M^{\prime}N^{\prime}}(g^{\prime\prime})^*\;
\psi(g^{\prime\prime}).
\end{eqnarray}
This involves a group element
$s(g^{\prime},g^{\prime\prime})\;\in \;G$, the `mid-point' of the group
elements $g$ and $g^\prime$, which is required to satisfy the following
conditions:
\begin{eqnarray}\label{7}
g^{\prime},
g^{\prime\prime}\;\in \;G &\rightarrow&
s(g^{\prime},g^{\prime\prime})\;\in \; G ,\nonumber\\
s(g^{\prime}, g^{\prime\prime})&=& s(g^{\prime\prime}, g^{\prime})
,
\nonumber\\ s(g^{\prime}, g^{\prime}) &=& g^{\prime} ,\nonumber\\
s\left(g_1 g^{\prime} g^{-1}_2 , g_1 g^{\prime\prime}
g^{-1}_2\right) &=& g_1 \;s(g^{\prime}, g^{\prime\prime}) g_2^{-1}
.
\end{eqnarray}
By virtue of $(\ref{7})$ and  $(\ref{2a}-\ref{2c})$ one finds that the Wigner
distribution $W(g;JMN\;M^{\prime}N^{\prime})$ corresponding to any
$\psi(g) \in {\cal H}$ possesses the following
properties:
\begin{itemize}
\item {\it Hermiticity}:
\begin{equation}
 W(g;JMN\;M^{\prime}N^{\prime})^* =W(g;JM^{\prime}N^{\prime}\;MN).
\end{equation}
\item {\it Marginals}
\begin{eqnarray}
\sum_{g\in G} \;W(g; JMN\;M^\prime N^\prime) &=& \psi_{JM^\prime N^\prime}
{\psi_{JMN}}^* , \nonumber\\
\sum\limits_{JMN} W(g; JMN\;MN) &=& |\psi(g)|^2 .
\end{eqnarray}
\item {\it Transformation under left translations}
\begin{eqnarray}
\psi^{\prime}(g) =\psi(g^{-1}_1 g)\rightarrow &&\nonumber\\
W^{\prime}(g;JMN\;M^{\prime}N^{\prime}) &=&\sum\limits_{M_{1}M_{1}
^{\prime}} {\cal D}^J_{MM_{1}}(g_1) {\cal D}^J_{M^{\prime}M^{\prime}_{1}}
(g_1)^* W\left(g^{-1}_1 g; JM_{1}N\;M^{\prime}_1
N^{\prime}\right).
\end{eqnarray}
\item {\it Transformation under right translations}
\begin{eqnarray}
\psi^{\prime\prime}(g) = \psi(g g_2)\rightarrow &&\nonumber\\
W^{\prime\prime}(g; JMN\;M^{\prime}N^{\prime})&=&
\sum\limits_{N_1 N_1^{\prime}} W\left(g g_2; JMN_1 M^{\prime}
N^{\prime}_1\right) {\cal D}^J_{N_1 N}\left(g_2^{-1}\right)
{\cal D}^J_{N^{\prime}_1 N^{\prime}}\left(g^{-1}_2\right)^*.
\end{eqnarray}
\item {\it Traciality}
If ${\hat \rho}_1$ and ${\hat \rho}_2$ denote two density
operators and $W_1(g;JMN\;M^\prime N^\prime)$ and
$W_2(g;JMN\;M^\prime N^\prime)$ the corresponding Wigner
distributions, then
\begin{equation}
\sum_{JMM^\prime} \frac{|G|}{N_J}\sum_{g\in G} {\tilde W}_1(g;
JMM^\prime) {\tilde W}_2(g; JM^\prime M)={\rm Tr}({\hat
\rho}_1{\hat \rho}_2), \label{78a}
\end{equation}
where the auxiliary function ${\tilde W}(g;JMM^\prime)$ appearing in
this equation is obtained from $W(g;JMN\;M^\prime N^\prime)$ by
setting $N=N^\prime$ and summing over $N$:
\begin{equation}
{\tilde W}(g;JMM^\prime)=\sum_N W(g;JMN\;M^\prime N).
\end{equation}
Since any density operator ${\hat \rho}_1$ is fully determined by
the traces of its products with all other density operators ${\hat
\rho}_2$, we can see that even the simpler function ${\tilde W}(g;
JMM^\prime)$ fully characterises ${\hat \rho}$ leading to the
conclusion that $W(g;JMN\;M^\prime N^\prime)$ captures information
contained in ${\hat \rho}$ in an overcomplete manner. This
overcompleteness, however, disappears in the Abelian case.
\end{itemize}
\noindent Note that any choice of a function $s(g^{\prime},
g^{\prime\prime})$ obeying conditions $(\ref{7})$  leads to an
acceptable definition of a Wigner distribution for quantum
mechanics on a group $G$.

The covariance conditions in the  last line in $(\ref{7})$ help us simplify
the problem of finding $s(g,g^\prime)$ to the choice of a suitable
function $s_0(g)$, the `square root' of the group element $g$,
 obeying the following conditions that ensure $(\ref{7})$:
\begin{eqnarray}\label{13}
 s(e,g) &=& s_0(g) ,\nonumber\\
s(g^{\prime},g^{\prime\prime})&=& g^{\prime} \;s_0(g^{\prime -
1}g^{\prime\prime}) :\nonumber\\ s_0(e) &=& e ;\nonumber\\
s_0(g^{-1}) &=& g^{-1} s_0(g) ;\nonumber\\ s_0(g^{\prime} g
g^{\prime -1}) &=& g^{\prime}\;s_0(g) g^{\prime -1} .
\end{eqnarray}
\noindent
A consequence of these conditions on $s_0(g)$ is that
\begin{equation}
s_0\;(g)\; g = g\;s_0\;(g) .
\end{equation}

Thus the problem of setting up a Wigner distribution for any finite
group reduces to constructing a function $s_0(g)$ for each $g\in
G$ satisfying $(\ref{13})$. This is easily accomplished if $|G|$
is odd. It is well known that for any finite group $G$ of odd
order the map $g\rightarrow g^2$ is one to one and onto. In other
words for every group element $g_i \in G $ there is a unique $g_k \in
G$ such that $g_i=g_{k}^2$. This being the case we can take
$s_0(g_i)= g_k$. ( One way of arriving at this result is to look
at the cycles generated by individual group elements. On the one
hand they are like one parameter subgroups or geodesics through
the identity in the Lie group case; on the other hand each $g$
obeys $g^{2m-1}=e$ or $g^{2m}=g$ for a least positive $m$ giving
$g^m$ for the square root of $g$) It is easily verified that such
a choice satisfies all the properties required of $s_0(g)$. This
in turn enables us to construct $s(g,g^\prime)$ for all pairs
$g,g^\prime \in G$ satisfying $(\ref{7})$ required for setting up
Wigner distributions for all finite groups of odd order.

 In the case of Abelian groups, as noted earlier, simplifications occur owing 
to the fact that all irreducible representations
 are one dimensional. Thus, for  the cyclic
 group ${\cal Z}_N=\{0,1,\cdots , N-1\}$ with $N$ odd, the formalism
 presented above, with appropriate notational changes, leads to
\begin{equation}
W(k,J)= \frac{1}{N}\sum_{l=0}^{N-1}\sum_{m=0}^{N-1}
\delta(k,s(l,m)) \psi^*(l){\cal D}^{J}(l)\psi(m){\cal D}^{J}(m)^*
~~;k,~J= 0,1,2,\cdots, N-1,
\label{16a}
\end{equation}
where
\begin{equation}
{\cal D}^{J}(k) = \omega^{kJ};~ \omega=\exp(2\pi i/N). \label{16}
\end{equation}
The mid point $ s(l,m)$  of two group elements $l,m$ is $n$ where
$2n = l+m~({\rm mod}~ N)$ and hence we can rewrite $(\ref{16a})$ as
\begin{equation}
W(k,J)= \frac{1}{N}\sum_{l=0}^{N-1} \psi^*(l)\psi(2k-l)
~\omega^{2J(l-k)}.
\end{equation}
It is gratifying to see this result, extensively discussed in the
literature \cite{10}~\cite{11}, come out from the general result in
$(\ref{10})$.

Next we turn to non Abelian finite groups. In particular we consider, in some 
detail, the smallest such group of odd order, a group of order $21$ consisting
of a semi-direct product $G={\cal Z}_7 \times {\cal Z}_3 $ of ${\cal Z}_7$ and 
${\cal Z}_3$. We can display $G$ as arising out of products of powers of two 
primitive generating elements $a$ and $b$ obeying the algebraic relations 
\begin{equation}
a^7~ =~ b^3 ~= ~e,~~~~b~a=a^2~b, 
\label{20n}
\end{equation}
where $e$ is the identity element of $G$. Thus $a$ and $b$ generate 
${\cal Z}_7$ and ${\cal Z}_3$ respectively. From $(\ref{20n})$ we derive :
\begin{eqnarray}
 b^j~a^\lambda &=& a^{2^j\lambda}~b^j,~~~j,\;\lambda \geq 0; \nonumber\\
b~a~b^{-1} &=& a^{2},~~~~b^{-1}~a~b=a^4, \nonumber\\
a~b~a^{-1} &=& a^{6}~b,~~~~a^{-1}~b~a=a~b. 
\label{21n}
\end{eqnarray}
We write the elements of $G$ as ordered products of a power of $a$ followed by
a power of $b$, and denote them by the corresponding pair of non negative
integers : 
\begin{eqnarray}
(\lambda,~j) &=& a^\lambda~b^j,~~~ 0~\leq~\lambda~\leq~6,~ 
0~\leq~j~\leq~2,\nonumber\\
(0,0)&=& e. 
\end{eqnarray}
The composition law then follows from $(\ref{21n})$,
\begin{equation}
(\lambda,~j)(\mu,~k) =(\lambda+2^j\mu~ {\rm mod}~7, ~j+k~{\rm mod}~3).
\label{23n}
\end{equation}
The elements $\{(\lambda,~0)\},~\{(0,~j)\}$ constitute the subgroups 
${\cal Z}_7,~{\cal Z}_3$ respectively. The semidirect product structure is 
evident, with ${\cal Z}_7$ being the invariant subgroup. 

Repeated use of $(\ref{21n})$ shows that the equivalence ( or conjugacy)
classes are five in number. Here we can use the fact that the exponent of 
$b$ is unchanged upon conjugation :
\begin{equation}
g\in G ~~:~~ g(\lambda,~j)g^{-1}=(\lambda^\prime,~j). 
\end{equation}
The classes are   
\begin{eqnarray}
{\cal C}_1 &=& \{(0,0)=e\};\nonumber\\
{\cal C}_2 &=& \{(\lambda~,0)=a^\lambda,~\lambda=1,~2,~4\};\nonumber\\
{\cal C}_3 &=& \{(\lambda,~0)=a^\lambda,~\lambda=3,~5,~6\};\nonumber\\
{\cal C}_4 &=& \{(\lambda,~1)=a^\lambda~b,~0~\leq~\lambda~\leq~6 \};\nonumber\\
{\cal C}_5 &=& \{(\lambda,~2)=a^\lambda~b^2,~0~\leq~\lambda~\leq~6 \}.
\label{25n}
\end{eqnarray}
There are therefore five inequivalent irreducible (unitary) representations 
of $G$. Since the squares of their dimensions must add up to $21$, we see that
there are three distinct one dimensional representations, and two distinct
three-dimensional ones. We label them by $J=1,~2,~3,~4,~5$ with dimensions 
$N_1=N_2=N_3 =1,~N_4=N_5=3$. 

The one dimensional representations obtain when the irreducible
representations of ${\cal Z}_3$ are simply lifted to $G$ and the element $a$
is realised trivially: 
\begin{eqnarray}
J=1,~2,~3 ~:~~ a\rightarrow 1,~~~~b&\rightarrow & \omega^{J-1},~~~\omega= 
\exp(2 \pi i/3)\nonumber\\
{\cal D}_{11}^{J}((\lambda,~j)) &=& \omega^{j(J-1)}.
\label{26n}
\end{eqnarray}

In the three dimensional representations we may assume that the matrix
representing $a$ is diagonal, with the eigenvalues being selected seventh
roots of unity. We also have from $(\ref{25n})$ the matrices representing 
$a,~a^2$ and $a^4$ (similarly $a^3,~a^5$ and $a^6$) are similarity transforms
of one another. One then easily arrives at the construction for, say, J=4:
\begin{eqnarray}
J=4~: ~~~&&a\rightarrow \pmatrix{\omega^\prime&0&0\cr
0&{\omega^\prime}^4&0\cr0&0&{\omega^\prime}^2\cr},~~~b\rightarrow 
\pmatrix{0&0&1\cr
1&0&0\cr 0&1&0\cr}, \nonumber\\
\nonumber\\
&&\omega^\prime = \exp(2\pi i/7). 
\label{27n}
\end{eqnarray}  
The primitive algebraic relations $(\ref{20n})$ are obeyed, so we do have a
representation of $G$. Irreducibility follows from Schur's lemma : since $a$
is diagonal non degenerate, any matrix commuting with it must be diagonal. If
it also commutes with the permutation matrix $b$, it must be a multiple of 
the identity. To calculate the elements of the representation matrices we
write $(\ref{27n})$, using ${\omega^\prime}^2={\omega^\prime}^9$, as 
\begin{eqnarray}
a_{MN} &=& \delta_{MN}~ (\omega^\prime)^{M^2}, \nonumber\\
b_{MN} &=& \delta_{[M+2],N}~,~~M,N=1,~2,~3~~, 
\end{eqnarray}
where $[M+2]$ indicates value ${\rm mod}~3$, in the range $1,~2,~3$. For
powers of $a$ and $b$ we then have : 
\begin{eqnarray}
(a^\lambda)_{MN} &=& \delta_{MN} ~(\omega^\prime)^{\lambda M^2}, \nonumber\\
(b^j)_{MN} &=& \delta_{[M+2j],N}~,    
\end{eqnarray}
leading to 
\begin{eqnarray}
{\cal D}_{MN}^{4}((\lambda,~j)) &=& (a^\lambda~b^j)_{MN}\nonumber\\ 
&=&(\omega^\prime)^{\lambda M^2}~\delta_{[M+2j],N}~~.
\label{30n}
\end{eqnarray}
The other three dimensional irreducible representation $J=5$ is the complex
conjugate of $J=4$. 
\begin{eqnarray}
J=5~:~~~&a&\rightarrow  \pmatrix{{\omega^\prime}^6&0&0\cr
0&{\omega^\prime}^3&0\cr0&0&{\omega^\prime}^5\cr},~~~b\rightarrow 
\pmatrix{0&0&1\cr
1&0&0\cr 0&1&0\cr}; \nonumber\\
\nonumber\\
(a^\lambda)_{MN} &=& \delta_{MN} ~(\omega^\prime)^{-\lambda M^2},
\nonumber\\
\nonumber\\
(b^j)_{MN} &=& \delta_{[M+2j],N}~~;\nonumber\\
\nonumber\\
{\cal D}_{MN}^{5}((\lambda,~j)) &=& 
(\omega^\prime)^{-\lambda M^2}~\delta_{[M+2j],N}~~.
\label{31n}
\end{eqnarray}

We next need expressions for the `square root' elements $s_0(g)$ and the 
`mid point' elements $s(g^\prime,g^{\prime\prime})$ for general 
$g,~g^\prime,~g^{\prime\prime} \in G$. For the former, from the composition
rule $(\ref{23n})$ we see that the square  of the element $(\mu,~k)$ is
given by 
\begin{eqnarray}
(\mu,~k)^2&=& (\lambda,~j)~,\nonumber\\
\lambda &=& (1+2^k)\mu~{\rm mod}~7,~~j= 2k~ {\rm mod}~3.
\end{eqnarray}

It is now indeed possible to solve for $\mu$ and $k$ uniquely in terms of 
$\lambda$ and $j$. By straight forward enumeration we find : 
\begin{eqnarray}
s_0((\lambda,~j)) = ((4+2^{2+j}-2^{2-j})&&\lambda ~{\rm mod}~7,~ 2j~
{\rm mod}~3 ), \nonumber\\
&& 0~\leq\lambda~\leq~6,~~0~\leq j~\leq~2.
\end{eqnarray}
As for the mid point element $s(g^\prime,g^{\prime\prime})$ we have from 
$(\ref{13})$ : 
\begin{equation}
s((\lambda,~j),~(\mu,~k)) =(\lambda,~j)~s_0((\lambda,~j)^{-1}(\mu,~k)).
\end{equation}
Since again from $(\ref{23n})$ we have 
\begin{eqnarray}
(\lambda,~j)^{-1}= (-2^{3-j}\lambda ~{\rm mod}~7,~&&2j~{\rm mod}~3),\nonumber
\\
&&0~\leq\lambda~\leq~6,~~0~\leq j~\leq~2,
\end{eqnarray}
the computation of $s(g^\prime,g^{\prime\prime})$ is complete. 
  
Armed with the knowledge of the mid point of two group elements
and the matrices of the irreducible representations of this group,
we can now compute the Wigner distribution for any state $\psi(g)$ 
belonging to the Hilbert space of complex valued functions on this 
non-Abelian group. Thus, for instance, for $\psi(g)$ given by 
\begin{equation}
\psi(g) = c_1 \delta_{g,(3,1)}+ c_2 \delta_{g,(2,2)};~~ |c_1|^2+|c_2|^2=1, 
\end{equation}
the Wigner distribution turns out to be  
\begin{eqnarray} 
W(g;JMN\;M^{\prime}N^{\prime})&=& |c_1|^2 W_1(g;JMN\;M^{\prime}N^{\prime})+
|c_2|^2 W_2(g;JMN\;M^{\prime}N^{\prime})\nonumber\\
&+&
W_{\rm int}(g;JMN\;M^{\prime}N^{\prime}),
\label{39}
\end{eqnarray}
where
\begin{eqnarray}
 W_1(g;JMN\;M^{\prime}N^{\prime})&=& \frac{N_J}{21} \delta_{g,(3,1)}
D^{J}_{MN}((3,1))D^{J}_{M^\prime N^\prime}((3,1))^*,\\
W_2(g;JMN\;M^{\prime}N^{\prime})&=& \frac{N_J}{21} \delta_{g,(2,2)}
D^{J}_{MN}((2,2))D^{J}_{M^\prime N^\prime}((2,2))^*,
\end{eqnarray}
and
\begin{eqnarray}
W_{\rm int}(g;JMN\;M^{\prime}N^{\prime})=\frac{N_J}{21} \delta_{g,(0,0)}[c_{2}^{*} c_1&& D^{J}_{MN}((2,2))D^{J}_{M^\prime N^\prime}((3,1))^*
\nonumber\\
&&+c_{1}^{*} c_2 ~D^{J}_{MN}((3,1))D^{J}_{M^\prime N^\prime}((2,2))^*].
\end{eqnarray}
The first two terms in $(\ref{39})$ contain the Wigner functions of wave
 functions $\psi_1(g)= \delta_{g,(3,1)}$ and $\psi_2(g)= \delta_{g,(2,2)}$. 
The third term contains the  `interference' effects  arising from the fact 
that the state $\psi$ is a superposition of $\psi_1$ and $\psi_2$. Owing to 
the specific structure of $\psi_1$ and $\psi_2$, while 
$W_1(g;JMN\;M^{\prime}N^{\prime})$ and $W_2(g;JMN\;M^{\prime}N^{\prime})$ have 
supports at $g=(3,1)$ and $g=(2,2)$ respectively, 
$W_{\rm int}(g;JMN\;M^{\prime}N^{\prime})$ has support at 
$g=(0,0)$, the mid point of $(3,1)$ and $(2,2)$. The values of 
$W_1(g;JMN\;M^{\prime}N^{\prime}), W_2(g;JMN\;M^{\prime}N^{\prime})$ and 
$W_{\rm int}(g;JMN\;M^{\prime}N^{\prime})$ can be explicitly computed using 
$(\ref{26n}),(\ref{30n})$ and $(\ref{31n})$. Thus, for instance, 
\begin{equation}
W_{\rm int}(g;412\;13)=\delta_{g,(0,0)}\times 
 \frac{1}{7}c_2^* c_1 \exp {\{-i\frac{8\pi}{7}\}}.
\end{equation} 
 
 To conclude, Fourier transforms on non Abelian finite
groups find a wide variety of applications in areas such as signal
processing, cryptology and quantum computation \cite{15},\cite{16}
and we hope that the Wigner distribution formalism
 for all finite groups of odd order developed here may provide an alternative
useful framework for examining some of these problems.


\begin{references}
\bibitem{1} E. P. Wigner, Phys. Rev. {\bf 40}, 749 (1932).
\bibitem{2} For a
comprehensive review see M. Hillery, R. F. O'Connell, M. O. Scully and
E. P. Wigner, Phys. Rep. {\bf 106}, 121 (1984) and also Y. S. Kim and
M. E. Noz, {\it Phase Space Picture of Quantum Mechanics}, (World Scientific,
Singapore, 1991); W. P. Schleich {\it Quantum Optics in Phase Space},
(Wiley-VCH, Weinheim, 2001);
R. L. Stratonovich, Zh. Eksp. Teor. Fiz. {\bf 31}, 1012 (1956) 
(Engl. Transl. Sov. Phys.-JETP, {\bf 4}, 891 (1957) ); 
 D. A. Dubin, M. A. Hennings and T. B. Smith, {\it Mathematical
    aspects of Weyl quantization and Phase}, (World Scientific, Singapore, 
2000).
\bibitem{3} G. S. Agarwal, Phys. Rev. A {\bf 24}, 2889 (1981);
 G. S. Agarwal, Phys. Rev. A {\bf 47}, 4608 (1993);
J. P. Dowling, G. S. Agarwal and W. P. Schleich, Phys. Rev. A {\bf 49},
4101 (1994).
\bibitem{4}J. C. V\'arilly and J. M. Gracia-Bond\'ia, Ann. Phys. NY,
{\bf 190} 107 (1989).
\bibitem{5} K. B. Wolf, Opt Commun. {\bf 132}, 343 (1996).
\bibitem{6} D. M. Kaplan and G. C. Summerfield, Phys. Rev. {\bf 187},
639 (1969).
\bibitem{7} C. Fronsdal, Rep. Math. Phys. {\bf 15}, 111 (1979).
\bibitem{8} C. Moreno and P. Ortega-Navarro, Lett. Math. Phys. {\bf 7},
 181 (1983).
\bibitem{9} R. Gilmore, {\it Lecture Notes in Physics}, {\bf 278}, ed.
Y. S. Kim and W. W. Zachary,  (Springer, Berlin 1987), p 211; W-M. Zhang,
D. H. Feng and R. Gilmore, Rev. Mod. Phys. {\bf 62}, 867 (1990).
\bibitem{10} W. K. Wootters, Ann. Phys. NY {\bf 176}, 1 (1987).
\bibitem{11} J.H. Hannay and M. V. Berry, Physica {\bf 1D},267 (1980);
 U. Leonhardt, Phys. Rev. Lett. {\bf 74}, 4101 (1995);  U. Leonhardt,
Phys. Rev. A {\bf 53}, 2998 (1996); C. Miguel, J. P. Paz, and M. Saraceno,
quant-ph/0204149,; j. P. Paz,quant-ph/0204150 .
\bibitem{12} C. Brif and A. Mann, J. Phys. A {\bf 31}, L9 (1998);
Phys.Rev. A {\bf 59}, 971 (1999).
\bibitem{13} N. Mukunda, Arvind, S. Chaturvedi and R. Simon, 
quant-ph/0305012. To appear in J. Math. Phys.~.
\bibitem{14} N. Mukunda, Am. J. Phys. {\bf 47}, 182 (1979).
\bibitem{15}  A. Terras, {\it Fourier Analysis on Finite Groups and
Applications} (Cambridge Univ. Press, 1999).
\bibitem{16} R. Jozsa quant-ph/0012084.
\end{references}
\end{document}